\documentclass[a4paper]{jpconf}
\usepackage{psfig}
\usepackage{amsmath}
\usepackage{amssymb}
\usepackage{graphicx}

\usepackage{psfig}
\usepackage{amsmath}
\usepackage{amssymb}
\usepackage{graphicx}
\usepackage{color}
\usepackage[colorlinks=false]{hyperref}
\usepackage{color}
\hypersetup{
    colorlinks=false,
    pdfborder={0 0 0},
}
\usepackage{epstopdf}

\newcommand{\note}[1]{\emph{\textcolor{red}{}}}

\newcommand{\Ni}{{\ensuremath{^{56}\mathrm{Ni}}}}

\newcommand{\Ox}{{\ensuremath{^{16}\mathrm{O}}}}

\newcommand{\Cx}{{\ensuremath{^{12}\mathrm{C}}}}

\newcommand{\Ms}{{\ensuremath{{M}_{\odot} }}}

\newcommand{\erg}{{\ensuremath{\mathrm{erg}}}}

\newcommand{\CASTRO}{\texttt{CASTRO}}
\newcommand{\KEPLER}{\texttt{KEPLER}}

\renewcommand{\thefootnote}{\ifcase\value{footnote}\or*\or
(**)\or(***)\or(****)\or(\#)\or(\#\#)\or(\#\#\#)\or(\#\#\#\#)\or($\infty$)\fi}

\newcommand{\araa}{ARA\&A}%
\newcommand{\apj}{ApJ}%
\newcommand{\apjl}{{ApJ}}%
\newcommand{\apjs}{{ApJS}}%
\newcommand{\aap}{{A\&A}}%
\newcommand{\aapr}{{A\&A~Rev.}}%
\newcommand{\mnras}{{MNRAS}}%
\newcommand{\nat}{{Nature}}%
\makeatletter

\newcommand{\Rmnum}[1]{\expandafter\@slowromancap\romannumeral #1@}
\makeatother

\begin{document}
\title{Fluid Instabilities of Magnetar-Powered Supernovae}

\author{Ke-Jung Chen}

\address{Division of Theoretical Astronomy, National Astronomical Observatory of Japan, 
	Tokyo 181-8588, Japan \\ }

\ead{ken.chen@nao.ac.jp}
\begin{abstract}
Magnetar-powered supernova explosions are competitive models for explaining very luminous optical transits. 
Until recently, these explosion models were mainly calculated in 1D. Radiation emitted from the magnetar  
snowplows into the previous supernovae ejecta  and causes a nonphysical dense shell (spike) found in 
previous 1D studies. This suggests that strong fluid instabilities may have developed within the 
magnetar-powered supernovae. Such fluid instabilities emerge at the region where  luminous transits 
later occur, so they can affect the consequent observational signatures. We examine the magnetar-powered 
supernovae with 2D hydrodynamics simulations and find that the 1D dense shell transforms into 
the development of Rayleigh-Taylor and thin shell instabilities in 2D. The resulting mixing is able to fragment 
the entire shell and break the spherical symmetry of supernovae ejecta. 
\end{abstract}


\section{Introduction} 
Magnetars are neutrons with a strong magnetic field of $10^{14} - 10^{15}$ Gauss (G) and 
a rapid rotation of period about a few to tens of milliseconds. This strong magnetic field may be generated 
from the collapse of a rapidly rotating iron core \cite{duncan1992,thompson1993,wheeler2000,thompson2004}. 
The magnetic field acts as a brake for the rotating neutron stars by extracting its rotational energy 
through a dipole radiation. During the early evolution of a magnetar, the energetic radiation appears in 
the form of x-rays and soft gamma-rays, which have been detected \cite{woosley2006, kouv1998, gaensler2005, maeda2007, mereghetti2008,esposito2009}. 
In addition, recent studies by Woosley and Kasen\cite{woosley2010, kasen2010}  suggested that magnetars might be able to power 
supernovae (SNe) with luminous optical transits of $10^{44} - 10^{45} \erg$ s$^{-1}$, which is about 10 - 100 times more 
luminous than that of a generic core-collapse SN.  Woosley\cite{woosley2010} modeled the magnetar in one dimension (1D) 
with the \KEPLER{} code \cite{kepler,heger2001}.  In his simulations, a large density spike was found when the magnetar 
started to emit the dipole radiation. The spike was also seen in \cite{kasen2010}. The emerging density spike is originated 
from the growth of fluid instabilities, and 1D simulations cannot model fluid instabilities, which are intrinsically multidimensional 
phenomena.  It is not clear how mixing inside the magnetar  would change the observational signatures of the 
magnetars. However, the spectrum must be affected by the mixing of different elements within shells. 

3D radiation-hydrodynamical simulations are required to study the mixing of magnetar-powered SNe and to obtain 
the light curves and spectra from the first principles.  However, such simulations are still beyond the capability of 
state-of-the-art numerical codes and computational resources.  As a first step to this goal, we have performed 
the first 2D hydrodynamics simulations based on a realistic magnetar progenitor, but we neglect the full radiation 
transport. This setup is still effective because the  density spike emerges at the early phase of the magnetar when 
the radiation is still strongly coupled with the gas flow.   

The structure of the paper is as follows; in Section 2, we describe the progenitor model and 
the setup for 2D simulations.  In Sections 3 and 4, we present the results of simulations and discuss their 
astrophysical implications.  We conclude in Section 5.

\section{Magnetar Model and Numerical Method}
\begin{figure}[ht]
	\begin{center}
		\includegraphics[width=.8\columnwidth]{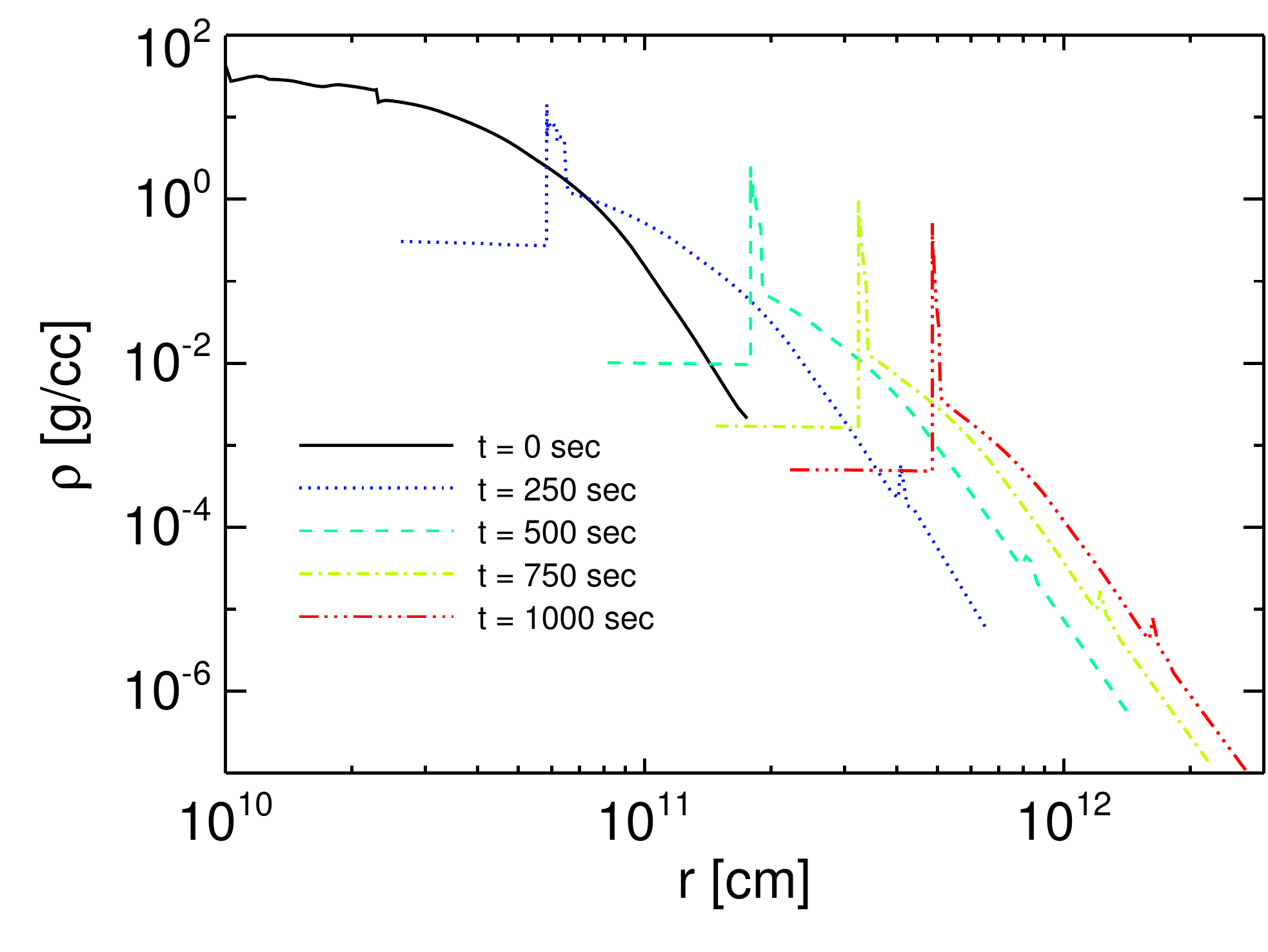}
		\caption[]{Density evolution of a magnetar-powered SN in 1D.  The curves
			show density at different times after the magnetar formation. A prominent 
			density spike emerges when the radiation-pressure-dominated gas snowplows 
			into previous supernovae ejecta. The amplitude of the spike grows very rapidly
			and exceeds a density contrast of $10^3$ within a thousand seconds.
  	\label{fig:rho}}
	\end{center}
\end{figure}

We start with the pre-supernova model of a  6 \Ms{}  carbon-oxygen star that has an initial mass 
fraction of $\Cx$ =0.144 and $\Ox$ = 0.856. The ratio is determined by abundances from the 
full-star stellar evolution models at the end of central helium burning. This particular model 
approximates the core of a regular star of $20 - 24\,\Ms$ with a solar metallicity based on the 
mass loss rate from \cite{jager1990}.  The relation between the CO core mass and the full star 
mass is still unclear due to the uncertainties of mass loss rate during stellar evolution. Mass loss 
may be driven by stellar wind, rotation, etc. Using a CO star allows us to directly study the evolution 
of pre-supernova by skipping its main sequence phase.  This model is evolved in \KEPLER{}, 
a 1D stellar evolution code, including hydrodynamics, nuclear burning and convection 
physics \cite{kepler}. The evolution is followed until an iron core of mass about $1.44$ \Ms{} 
forms when the core collapse is about to occur. The radius of the iron core is about 1,400 km, and 
it promptly collapses into a proto-neutron star about 12 km in size. The gravitational binding energy 
released from the iron-core collapse powers a supernova. The explosion mechanics behind the core-collapse 
SNe are still uncertain \cite{burrows1995,janka1996,mezz1998, murphy2008,nordhaus2010}. Hence, we 
use the piston explosion model by inserting kinetic energy of  $1.2\times 10^{51}$ erg at the outer edge of 
the iron core to blow up the star.  The explosion synthesizes about 0.22 \Ms\ $\Ni$ and creates a strong shock 
of velocity about $10^9$ cm s$^{-1}$ shortly after the explosion; we assume that a magnetar shortly forms 
with a rotation period of $1$ millisecond and a magnetic field stress of $4\times10^{14}$ G. This model produces 
a light curve that fits well with the observational data of a superluminous SN PTF10cwr \cite{Che16}. However, a 
prominent density spike forms again in \KEPLER\ and its amplitude density grows very rapidly. As shown in Figure~\ref{fig:rho}, 
the giant spike spanning over four orders of magnitude forms within 1,000 seconds.   1D Density spikes usually originate in the fluid instabilities, 
which 1D simulations are not capable of modeling. To address this issue, we now map the resulting 1D profiles into \CASTRO\ code
right before the emergent density spike and follow the simulation in 2D.

\subsection{2D \CASTRO{} Setup}
We run 2D simulations with \CASTRO, a multidimensional adaptive mesh refinement 
(AMR) hydrodynamics code \cite{ann2010, zhang2011}.  It uses an unsplit piecewise 
parabolic method (PPM) hydro scheme \cite{woodward1984} with multispecies advection.  
We use the helmoltz equation of state from \cite{timmes2000}, which considers the 
relativistic electron and positron pairs of arbitrary degeneracy, ions, which are treated as an ideal gas, 
and photons.  1D \KEPLER\ profiles of densities, velocities, temperatures, and different 
isotope abundances are mapped onto a 2D cylindrical grid of \CASTRO\ before the formation 
of the density spike.  The 1D-to-2D mapping is done with a scheme from Chen\cite{chen2013}, which 
conservatively maps the physical quantities such as mass and energy from 1D profiles onto 
multidimensional grids.

We simulate only an octant part of the star in 2D. The physical size of the domain  in $r$ and 
$z$ is $2\times$ 10$^{12}$ cm, which is about fifteen times larger than the radius of the progenitor
star. The circumstellar medium (CSM) is filled with an ambient gas of density profile $\rho \propto r^{-3.1}$
starting from the surface of the star. This density profile can prevent any artificial mixing caused by the 
reverse shock when the forward shock runs into the CSM. The base grid has $256\times256$ zones, with 
six levels of AMR for an additional factor of up to 64 (2$^6$) in spacial resolution. 
The grid refinement criteria are based on gradients of density, velocity, and pressure. 
The hierarchy nested grids are also constructed in such a way that the energy-deposited region is highly resolved. 
Reflecting and outflow boundary conditions are set on the inner and outer 
boundaries in both $r$ and $z$, respectively. We use the monopole approximation for self-gravity, 
in which a 1D profile of gravitational force is constructed from the radial average of the density, then the 
g-field stress of each grid is calculated by the linear interpolation of the 1D profile. This approximation is 
efficient and well-suited in supernova simulations.  A point source gravity of the magnetar is also taken 
into account. 

\begin{figure}[ht]
	\begin{center}
		\includegraphics[width=.8\columnwidth]{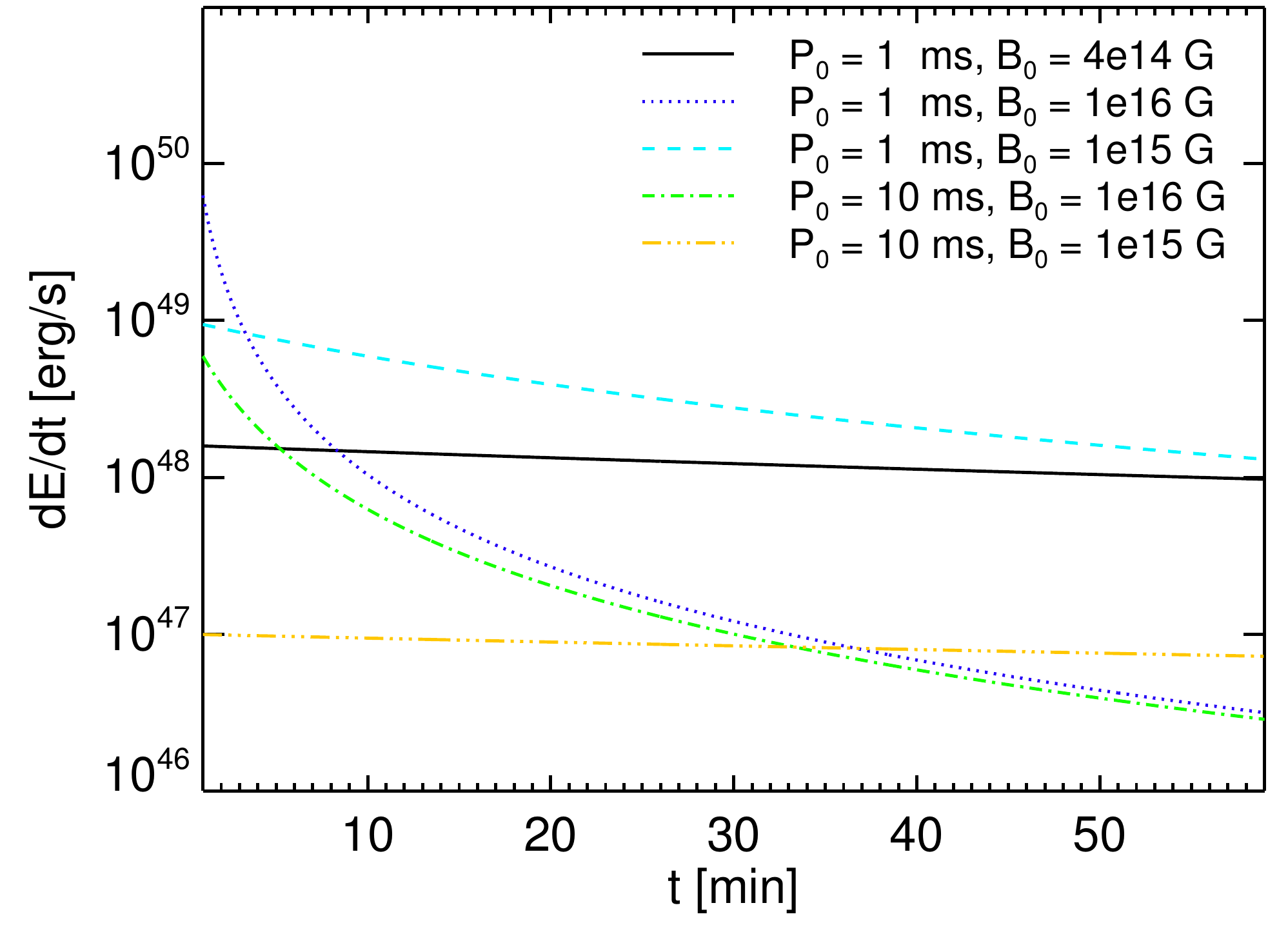}
		\caption[]{Energy generation rates of a magnetar. Curves show the 
			energy generation rates of magnetars with different 
			initial magnetic fields, B$_0$, and rotational periods, 
			P$_0$.  A Magnetar of $P =1$ millisecond and 
			$B_o = 4\times10^{14}$ G has a roughly constant rate 
			of about $10^{48} \erg$ s$^{-1}$ within its first hour.
			\label{fig:power}}
	\end{center}
\end{figure}

We assume that the magnetar releases its rotational energy through a dipole radiation. 
We use the moment of inertia  for a typical neutron star, $I\approx10^{45}$ g cm$^2$  
with a millisecond period, $P_{ms}$. Then, its rotational energy can be estimated as
\begin{equation}
E = \frac{1}{2}I\omega^2\approx 2\times 10^{52}P_{\rm ms}^{-2} \quad \erg;
\label{eq:1}
\end{equation}
$E$ is dissipated through a dipole radiation described by the Larmor 
formula \cite{lyne1990}. We assume the radius of a neutron star, $R_n\approx10^{6}$ cm, and
the inclination angle between the magnetic and rotational axes, $\alpha = 30^{\circ}$.   
The energy dissipation rate has the form
\begin{equation}
\begin{split}
\frac{dE}{dt} & = -\frac{32\pi^4}{3c^2} (BR_n^3\sin\alpha)^2P^{-4} \\
& \approx - 10^{49}B_{15}^2 P_{\rm ms}^{-4} \quad \erg.
\end{split}
\label{eq:2}
\end{equation}
Solving Equations~(\ref{eq:1}) and (\ref{eq:2}) by assuming the constant magnetic field,   
we can obtain  
\begin{equation}
P_{\rm ms}(t) \approx \sqrt{P_0^2+\frac{B_{15}^2t}{2000}}, 
\end{equation}
where $B_{15}= B/10^{15}$ G, $P_0 = P_{\rm ms}(0)$. We show the evolution of $E$ 
with different initial B-field and rotational rates in Figure~\ref{fig:power}. Briefly 
speaking, the initial amplitude of energy-release rates, $\epsilon = \dot{E}$, is 
determined by the initial rotational rate $\propto P_0^{-1}$, its decay rate, 
$\dot{\epsilon}$ is determined by $B_0$.  Strong $B_0$ acts like a strong brake 
that quickly extracts the rotational energy of the magnetar. In 2D \CASTRO\ simulations, 
the energy released from the dipole radiation is uniformly dumping within a sphere 
of  $r\approx 5 \times 10^{9}$ cm, which is about the location of the emergent 
spike seen in the 1D \KEPLER\ model.  
We deposited the magnetar energy in the small volume near the center of the star by injecting the thermal energy to the gas. 
At the same time, we added a very small amount of mass of  $2.5\times10^{-6}$ \Ms s$^{-1}$ along with the energy injection. Therefore, it 
can produce a high-velocity wind ($\sim 0.5c$) and does not violate the mass conservation of the simulation. Strictly speaking, the outflow comes from the magnetar should be highly magnetized and relativistic. But we do not know the  nature of 
the magnetar wind. In our current study, we focus on the dynamics of ejecta and fluid instabilities. It is reasonable that we assume 
the gas and radiation tightly couple at this early phase of magnetar evolution. So we directly dump the magnetar energy to tiny additional mass to form a wind. Once the ejecta becomes very optically thin, this assumption becomes invalid. We then need the radiation transportation simulations to follow the decoupled radiation and gas dynamics.

 The \CASTRO\ simulation is evolved until the radiation-pressure-dominated ejecta driven by the dipole radiation 
has expanded to $r \approx 2\times10^{12}$ cm, about 3,000 seconds after the magnetar forms.

\section{Results}

\begin{figure}[ht]
	\begin{center}
		\includegraphics[width=.8\columnwidth]{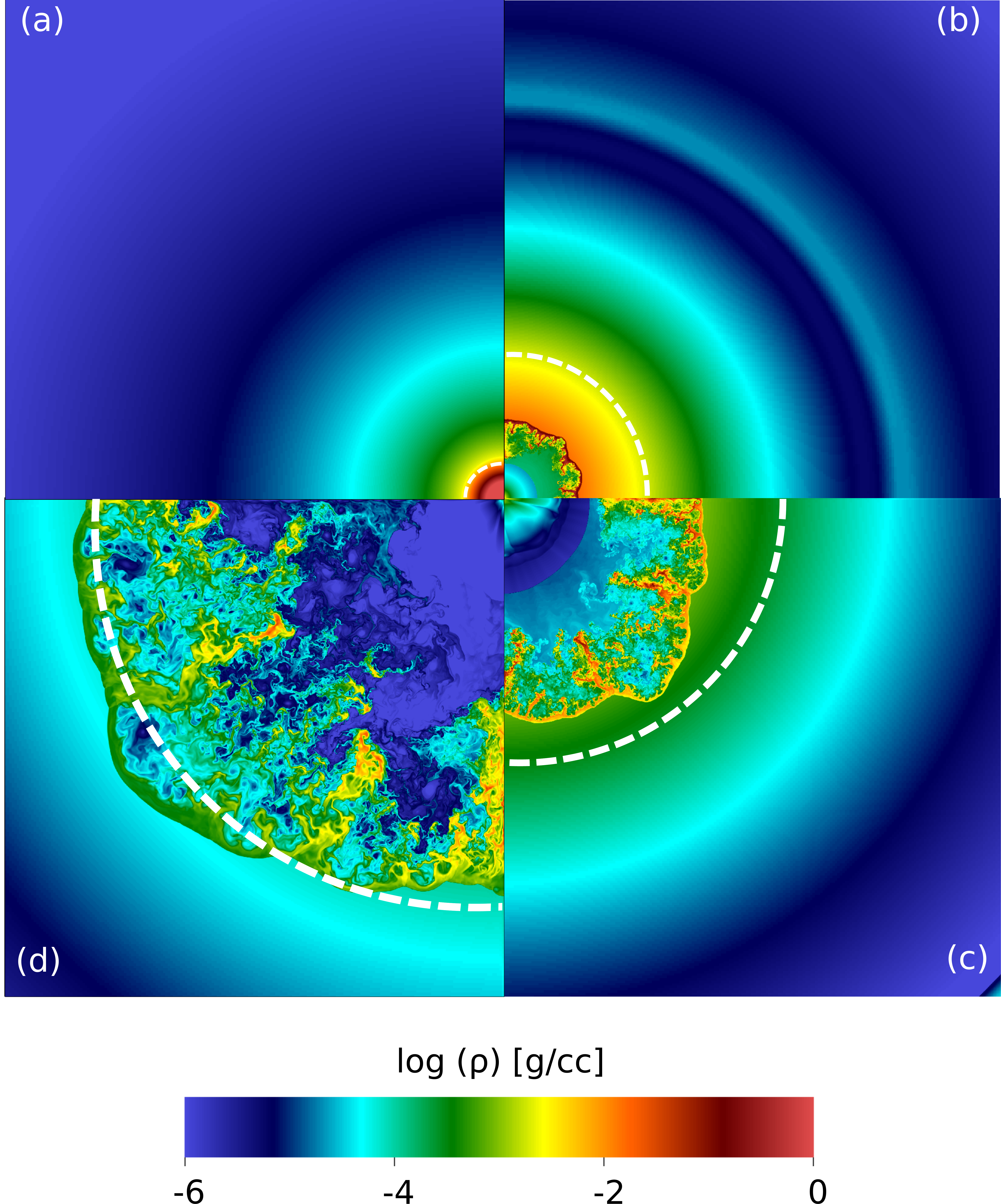}
		\caption[]{Evolution of fluid instabilities in the early phase 
			of 1 ms model. The white dashed-line arcs indicate 
			the outer boundary of expanding ejecta.  Panels (a) -- (d) show 
			densities at 0, 800, 1,600, and 2,400 seconds, 
			respectively. In Panel (a), the initial density shell 
			shows a spherical symmetry. After the magnetar forms, 
			fluid instabilities grow from tiny fingers to a large 
			scale mixing as shown in Panel (d). The shell of matter 
			close to the boundary of Panel (b) is caused by the shock 
			wave from the original supernova explosion.						
			\label{fig:2d1}}
	\end{center}
\end{figure}

\begin{figure}[ht]
	\begin{center}
		\includegraphics[width=.8\columnwidth]{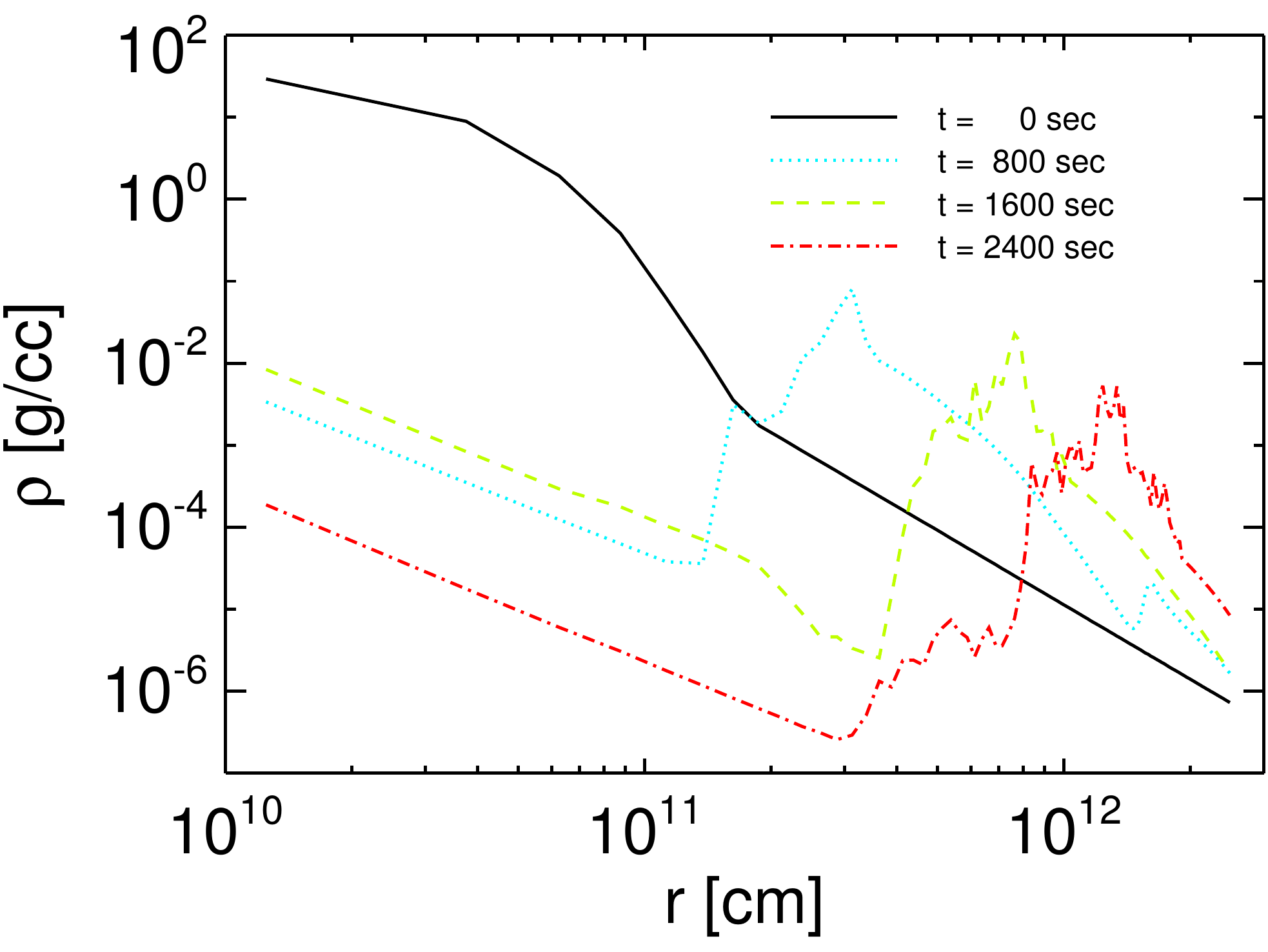}
		\caption[]{Angle-averaged density profiles of the 2D magnetar. Curves 
			represent angle-averaged profiles of snapshots shown 
			in Figure \ref{fig:2d1}. The fluid instabilities in 2D 
			truncate the 1D density spike into mixing.
			\label{fig:rho2d}}
	\end{center}
\end{figure}

The energy injection of the magnetar quickly heats up the surrounding gas to a temperature 
about $T\approx 5\times10^{8}$ K, apparently too low to ignite any interesting nuclear burning. 
Instead, the thermalized gas forms an outflow and runs into the previous SN ejecta. The gas driven 
by the outflow starts to pile up and form into a shell, and the region underneath the shell becomes 
radiation dominated. We define the entity of within the shell as a "radiative bubble". The bubble 
quickly expands at a rate of $2-5\times10^{8}$ cm/s. It takes about 600 seconds for the bubble to 
expand to the size of the progenitor star, $r_* \approx 2\times 10^{11}$ cm. The hot gas flow collides 
the shell, resulting in the fluid instabilities. In 1D \KEPLER\ models, there is no additional dimension to 
relax such collisions, and gas simply piles up to form a big density spike. The emergent fluid instabilities 
first form tiny little fingers that grow and fragment as the bubble expands. These fluid instabilities drill the spherical 
dense shell ahead of them and slightly break down the spherical symmetry of the bubble.  The fluid instabilities 
continue evolving as the bubble expands. We show the evolving fluid instabilities in Figure~\ref{fig:2d1}.  It is visible 
that the structure continue evolving as the bubble expands. Significant mixing has occurred inside the bubble and 
breaks down its spherical symmetry. In Panel (d) of Figure~\ref{fig:2d1},  mixing layers start to fulfill the bubble and 
mix up the isotopes when the bubble has reached to about eight times larger than $r_*$.  In Figure~\ref{fig:rho2d}, 
we show the angle-averaged profiles of density. The spikes seen in the 1D \KEPLER\ models disappear in 2D and 
result in a noisy bump, which is the site of mixing in action. Because 2D fluid is able to follow the mixing instead of 
piling up in a very thin shell,  it transforms the  density spike into mixing. The density constraint, $\delta \rho = {\rho-\langle\rho\rangle/\rho}$ , 
inside the mixing region is about 10 -- 100 instead of $10^3$ (spike) found in 1D models. In 1D models, most radiation is emitted 
from the density spike, which suggests the radiation may come from the mixing region in the multidimensional models.  

\begin{figure}[ht]
	\begin{center}
		\includegraphics[width=.8\columnwidth]{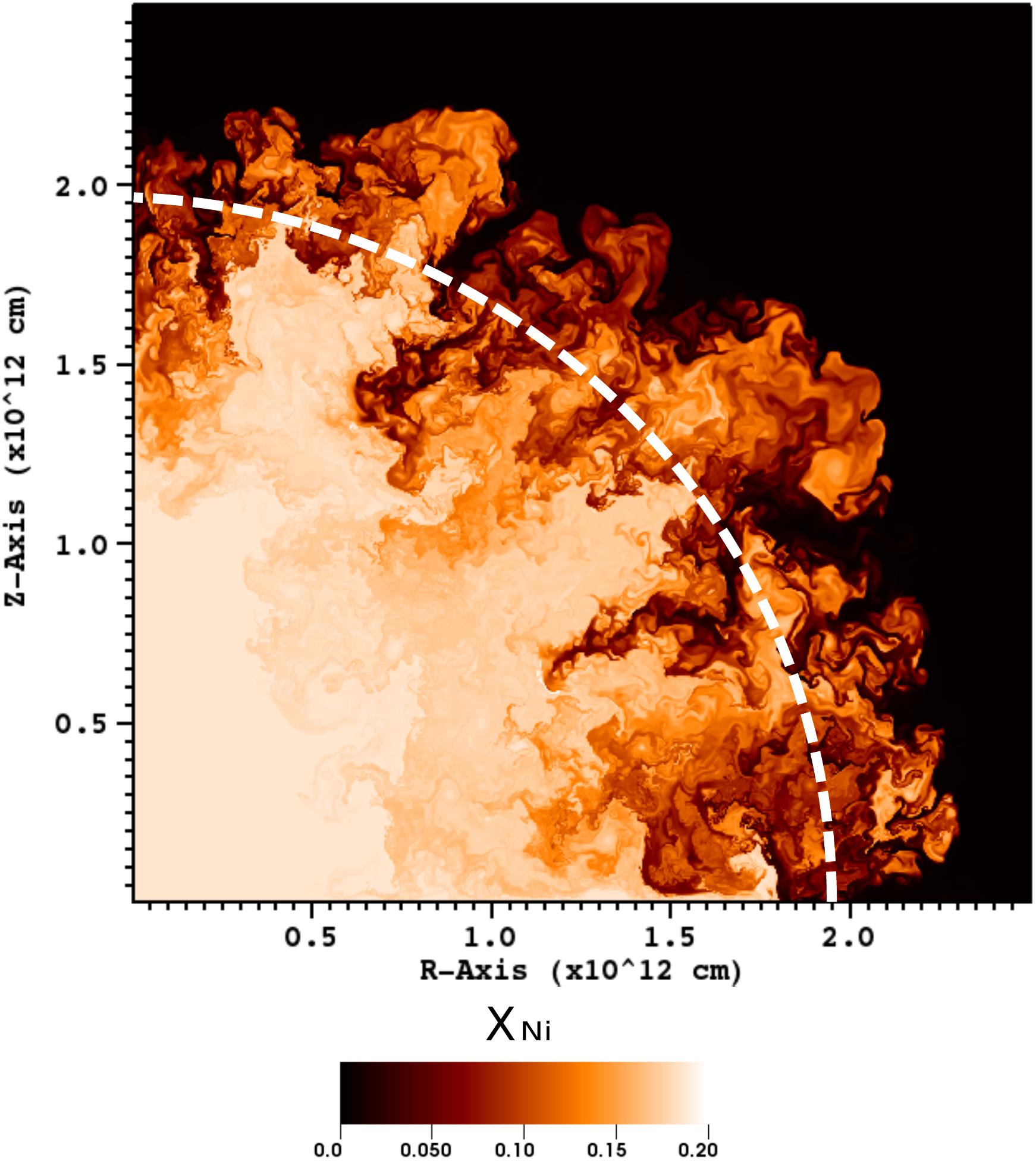}
		\caption[]{2D \Ni\ distribution at its final snapshot.  The white dashed-line arc indicates the outer boundary of
			 the expanding SN ejecta.\Ni\ is dredged up by the mixing, and some fraction of it starts to expand into the 
			 optical thin region. Energetic radiation directly from \Ni\  decay may be visible. \label{fig:ni}}
	\end{center}
\end{figure}

To examine the mixing, we show the mixing of \Ni\ at  end of the simulation in Figure~\ref{fig:ni}. Some fraction of \Ni{} appears at edge of 
the fragmented shell. If such dredging up of \Ni{} indeed happens at an early phase of magnetar evolution, there arises the possibility of 
gamma-ray emission from the \Ni{} decay.

\section{Discussion}
The dipole radiation heats the gas around the magnetar and drives a gas outflow. 
It creates a pressure gradient between the heated and unheated gas. However, the 
density gradient of gas is in a reverse direction of the pressure gradient. 
It fulfills the Rayleigh-Taylor (RT) instability criterion \cite{chan1961}. This creates the 
initial mixing under the shell, as we have seen. Once the shell expands, the shell propagates 
away from the magnetar, and it starts to experience  nonlinear thin shell instability (NTSI) 
\cite{vish1994}. NTSI is the strong driver of mixing and eventually breaks down the shells' 
spherical symmetry. Previous work\cite{chev1992,jun1998, blondin2001} have studied fluid instabilities 
in the context of pulsar wind nebula, which is related to our study.  

Dipole radiation may directly break out the thin dense layer of the radiative bubble, and escaping radiation can 
be detected in the form of hard x-ray emission. In addition, the mixing driven by the fluid instabilities has
altered the dynamics and chemical compositions of SN ejecta. Since mixing is strongest in the region of the 
flow from which most of the radiation originates, it likely affects the SN light curves and spectra.  There is also a 
possibility that some freshly synthesized \Ni{} about 0.03 \Ms\ can appear in the outskirts of the SN ejecta and can 
be examined in the supernova remnant. 

In our simulations, we do not consider the full radiation transport that later may become important in the 
magnetar-powered SNe because the radiative cooling of ejecta may affect its dynamics as well as its 
observational signatures. The earlier fragmentation of ejecta in our simulations may seed the large-scale 
inhomogeneity at an early time.  We also assume the constant $B$ field stress in this calculation; however provide $B$ field stress may 
decay in a spin down magnetar. The different parameters of $B$ and $P$ provided different energy injection rates that 
may affect the growth rates of fluid instabilities and change the  amount of escaping dipole radiation of magnetars.

\section{Conclusion}
We present realistic 2D hydrodynamical simulations of magnetar-powered supernovae.  
Due to the limits of dimensionality, previous 1D models cannot model the fluid instabilities and 
mixing from the first principles, thus it produces an unphysical density spike. Instead, our 2D 
simulations show that strong fluid instabilities and the resulting mixing dampens the formation 
of the density spike found in 1D. The results suggest that strong fluid instabilities 
occur in magnetar-powered supernovae. Similar results are expected to appear in 3D.  
These fluid instabilities are mainly driven by  dipole radiation of magnetar, which causes 
Rayleigh-Taylor instabilities and nonlinear thin shell instabilities. The resulting mixing 
transforms the supernova ejecta into filamentary structures. The morphology of mixing  looks 
similar to that seen in the Crab Nebula. It may suggest the Crab Nebula might already have 
formed filamentary structures very early on. The growth of fluid instabilities depends on the stellar 
structure and the physical properties of the magnetars, such as magnetic field stress and rotation rate. 
In this paper, we assume a constant magnetic field, which is a crude approximation. 
We neglect radiative cooling by metals, and dust, which rapidly cool the dense clumps when the 
ejecta become optically thin. This, in turn, can affect the growth rate of Rayleigh-Taylor instabilities.

Magnetar-powered supernovae  are competitive models for the superluminous 
supernovae, that possibly serve as promising probes for the early universe. 
It is crucial to obtain more realistic observational signatures of magnetar-powered 
supernovae, which will become more frequently observed in the coming supernova surveys. 
But they do demonstrate that multidimensional radiation transport will be required to 
model how photons are emitted from the complex structures caused by fluid instabilities.  
In future work, we will use the newly commissioned radiation 
hydrodynamics version of \CASTRO\ \cite{zhang2013} to better examine these 
explosions and to offer realistic observational diagnostics.

\section*{Acknowledgments}
I thank Stan Woosley and Tuguldur Sukhbold for many useful discussions. 
I acknowledge the support of the EACOA Fellowship from the East Asian Core Observatories Association. 
Work at UCSC has been supported by an IAU-Gruber Fellowship, the DOE HEP Program (DE-SC0010676), 
and the NASA Theory Program  (NNX14AH34G).  Numerical simulations are supported by the Minnesota 
Supercomputing Institute (MSI), the National Energy Research Scientific Computing Center (NERSC), and 
the Center for Computational Astrophysics (CfCA) at National Astronomical Observatory of Japan (NAOJ).

\newpage
\section*{References}


\clearpage

\end{document}